\documentclass[12pt,twoside,a4paper]{article}
\usepackage[dvips]{epsfig}
\newcommand{\gquarkS}{$\gamma^* q \rightarrow \gamma q g$}
\newcommand{\ggluonS}{$\gamma^* g \rightarrow \gamma q \bar{q}$}
\newcommand{\gonepone}{$\gamma + (1+1)$}
\newcommand{\gtwopone}{$\gamma + (2+1)$}
\newcommand{\gnpone}{$\gamma + (n+1)$}
\newcommand{\simgt}{\,\rlap{\lower 3.5 pt \hbox{$\mathchar \sim$}}
  \raise 1pt \hbox {$>$}\,}
\newcommand{\simlt}{\,\rlap{\lower 3.5 pt \hbox{$\mathchar \sim$}}
  \raise 1pt \hbox {$<$}\,}

\begin{document}
\thispagestyle{empty} 
\title{
\vskip-3cm
{\baselineskip14pt
\centerline{\normalsize DESY 99--084 \hfill ISSN 0418--9833}
\centerline{\normalsize MZ-TH/99--25 \hfill} 
\centerline{\normalsize TTP99--32 \hfill}
\centerline{\normalsize hep--ph/9907511 \hfill} 
\centerline{\normalsize July 1999 \hfill}} 
\vskip1.5cm
Photon Fragmentation in Large-$Q^2$ $ep$ Collisions \\
at Next-to-Leading Order QCD 
\author{A.~Gehrmann-De Ridder$^1$, G.~Kramer$^2$ and H.~Spiesberger$^3$
\vspace{2mm} \\
{\normalsize $^1$ Deutsches Elektronen-Synchrotron DESY,
D-22603 Hamburg, Germany, and} \\
{\normalsize Institut f\"ur Theoretische Teilchenphysik,
  Universit\"at Karlsruhe,}\\ 
{\normalsize D-76128 Karlsruhe, Germany} \vspace{2mm}\\
{\normalsize $^2$ II. Institut f\"ur Theoretische
  Physik\thanks{Supported by Bundesministerium f\"ur Forschung und
    Technologie, Bonn, Germany, under Contract 05~7~HH~92P~(0), and
    by EU Fourth Framework Program {\it Training and Mobility of
    Researchers} through Network {\it Quantum Chromodynamics and
    Deep Structure of Elementary Particles}
    under Contract FMRX--CT98--0194 (DG12 MIHT).}, Universit\"at
  Hamburg,}\\ 
\normalsize{Luruper Chaussee 149, D-22761 Hamburg, Germany} \vspace{2mm}
\\ 
\normalsize{$^3$ Institut f\"ur Physik,
  Johannes-Gutenberg-Universit\"at,}\\ 
\normalsize{Staudinger Weg 7, D-55099 Mainz, Germany} \vspace{2mm} \\
\normalsize{e-mail: gehra@particle.physik.uni-karlsruhe.de,} \\
\normalsize{kramer@mail.desy.de, hspiesb@thep.physik.uni-mainz.de}
} }

\date{}
\maketitle
\begin{abstract}
\medskip
\noindent
We study the production of photons accompanied by jets in large-$Q^2$
deep inelastic scattering. Numerical results for the cross section
differential with respect to the fraction of momentum $z_{\gamma}$
carried by a photon inside a jet at large $z_{\gamma}$, up to $O(\alpha
\alpha_s)$ in perturbative QCD, are presented.  The sensitivity to the
fragmentation contribution allows one to study the quark-to-photon
fragmentation function. Our results can be confronted with future
experimental data from HERA.
\end{abstract}

\clearpage
\section{Introduction}
The production of final state photons at large transverse momenta in
high-energy processes is an important observable for testing QCD. Data
on large-$p_T$ photon production in hadronic collisions have been used
in the past to obtain information on the gluon distribution in the
photon.  In addition, a good understanding of direct photon production
within the framework of the standard model will be essential for new
physics searches at future colliders.

Photons from high-energy collisions are produced essentially by two
mechanisms: the direct production of a photon off a primary quark or
antiquark, or the fragmentation of a hadronic jet into a single photon
carrying a large fraction of the jet energy. The direct production gives
rise to perturbatively calculable short-distance contributions whereas
the fragmentation contribution is due primarily to a long distance
process which can not be calculated completely inside perturbative QCD.
The latter is described by the process-independent quark-, antiquark-, 
or gluon-to-photon fragmentation functions (FF) \cite{1} which must be
fixed by experimental data as long as they can not be calculated by
non-perturbative methods. Their evolution with the factorization scale
$\mu_F$ and their so-called point-like contribution (up to a
normalization) can, however, be calculated perturbatively.

Directly produced photons are usually well separated from hadron jets,
while photons originating from the fragmentation process are primarily
found inside hadronic jets. By imposing an isolation criterion on the
photon, one can, in principle, suppress (but in general, not eliminate)
the fragmentation contribution to the final state photon cross
section\footnote{See \cite{2} for a proposal to eliminate the dependence
on the photon FF.}.

So far, only a limited number of measurements of single photon
production exists through which direct information on the photon FF can
be obtained.  A possible way is the measurement of the inclusive photon
cross section in $e^+e^-$ annihilation. Recently the OPAL collaboration
at LEP has measured this cross section for final state photons in the
range $0.2 < x_{\gamma} < 1.0$, where $x_{\gamma}=2E_{\gamma}/M_Z$ is
the photon energy fraction in terms of the beam energy \cite{3}.  The
results are in reasonable agreement with predictions obtained within
perturbative QCD using the model-dependent parametrizations of the
photon FF of DO \cite{6}, GRV \cite{4} or BFG \cite{5}, when choosing
the factorization scale $\mu_F=M_Z$.  The DO model for the FF is based
on an asymptotic solution of the evolution equation whereas the other
two models \cite{4,5} contain in addition a non-perturbative input
inspired by the vector meson dominance model.  Unfortunately, the
experimental precision was not sufficiently high to discriminate between
these theoretical predictions.  Note also that in \cite{10,12} it was
shown that predictions obtained within perturbative QCD and using a
fixed-order expanded expression for the photon FF also agreed well with
the OPAL data.

An alternative way to determine the FF from $e^+e^-$ data is to
measure the production of photons accompanied with a definite number of
hadronic jets \cite{7}. In this approach, the photon is treated like any
other hadron and is clustered simultaneously with the other hadrons into
jets (so-called democratic clustering procedure) \cite{8}. Then, one of
the jets in the final state is considered as the photon jet if the
fraction of the electromagnetic energy inside the jet is sufficiently
large, i.e.\ 
\begin{equation} 
  z = \frac{E_{\gamma}}{E_{\gamma}+E_{had}} > z_{cut}
\label{zzcut}
\end{equation}
with $z_{cut}$ fixed by the experimental conditions. On this basis the
ALEPH collaboration at LEP analysed events produced at the $Z$-resonance
\cite{9} which contained one hadron jet and a jet with a photon carrying
at least $70\%$ of the jet energy. This $\gamma + 1$-jet rate was used
to determine the quark-to-photon FF, which was calculated in leading
order (LO), i.e.\ up to O($\alpha$), in \cite{8}. The calculation was
extended to next-to-leading order (NLO), i.e.\ up to
O($\alpha\alpha_s$), in \cite{10,11} and a NLO photon-to-quark FF was
obtained \cite{10} by comparing with the ALEPH data.  Due to the
unfavorable signal-to-background ratio below $z=0.7$ for this
particular observable, the $\gamma + 1$-jet rate, the photon FF could be
determined only in the range $0.7 < z < 1.0$.  In \cite{12,13} also the
predictions of GRV \cite{4} and BFG \cite{5} for the photon FF were
compared to the ALEPH $\gamma + 1$-jet data. It turned out that the BFG
parametrization is in agreement with the measurements while the rate
predicted with the help of the FF of GRV lies too high.

The photon FF is process independent as any other FF and can be used to
predict the cross section for the production of single photons in other
processes. This has been done recently for isolated photon production in
deep-inelastic $ep$ scattering (DIS) \cite{26}. In that work the
emphasis was put on the prediction of cross sections in the HERA
kinematic range without the necessity to introduce photon-parton
separation cuts as used in earlier work \cite{27}.  This was achieved by
including the divergent photon-quark (-antiquark) singular contributions
and the bare quark-to-photon fragmentation contributions, leaving in the
final result only a finite factorization scale dependent quark-to-photon
FF contribution.  In our previous work we adopted as the photon FF the
lowest-order fit obtained from the photon+1-jet data of the ALEPH
collaboration \cite{9}.

In the present work we study the influence of the fragmentation
contribution on the cross section for the production of a photon plus
jets in DIS and give predictions for $d\sigma/dz_{\gamma}$ as a function
of $z_{\gamma}$ in the interval $z_{cut}=0.5<z_{\gamma}<1.0$ (for the
definition of $z_{\gamma}$ see below).  For this purpose we calculate
$d\sigma/dz_{\gamma}$ with the same cuts on the DIS variables as in
\cite{26} for essentially three different sets of photon FF's, the ALEPH
sets in LO and NLO \cite{9,10}, the leading-logarithmic (LL) and
beyond-leading logarithmic (BLL) parametrizations of GRV \cite{4} and
the BLL parametrization of BFG \cite{5}.  The first set of photon FF is
obtained within a fixed-order framework at a given order in
$\alpha_{s}$. The two other sets were obtained after the leading and/or
next-to-leading logarithms of the factorization scale $\mu_{F}$ were
resummed. The essence of these two approaches has been described
extensively in \cite{12} and will be outlined below.

\section{Leading and Next-to-Leading Order Cross Section}
In leading order\footnote{Here and in the following, we do not count the
  extra factor $\alpha$ from the $e-\gamma^*$ vertex.} (i.e., at
O($\alpha$)), the cross section for the production of photon plus jets
in DIS receives contributions from the quark (antiquark) subprocess
$\gamma^* q \rightarrow \gamma q$ ($\gamma^* \bar{q} \rightarrow \gamma
\bar{q}$). Together with the remnant jet from the proton this process
gives rise to a \gonepone-jet final state. In the virtual photon-proton
center-of-mass system the hard $\gamma$ recoils against the hard jet
back-to-back. Cuts on the usual DIS variables $x$, $y$ and $Q^2$ are
applied to remove $\gamma$ production by incoming $\gamma^*$'s of small
virtuality. To produce $\gamma$'s of sufficiently high energy, an
explicit cut on the total $\gamma^* p$ center-of-mass energy, $W$, is
introduced.  Both leptons and quarks emit photons. The leptonic
radiation is suppressed by a cut on the photon emission angle with
respect to the incoming electron in the same way as in the earlier work
\cite{26,27}.  In LO each parton is identified with a jet and the photon
is automatically isolated from the quark jet by demanding a non-zero
transverse momentum of the photon or jet in the $\gamma^* p$
center-of-mass frame. At this order there is no fragmentation
contribution.

At next-to-leading order (O($\alpha\alpha_s$)) we have subprocesses with
an additional gluon, either in the final or in the initial state, i.e.\ 
the subprocesses \gquarkS\ (and similarly with $q$ replaced by
$\bar{q}$) and \ggluonS, respectively. In addition, virtual corrections
(one-loop diagrams at O($\alpha\alpha_s$)) to the LO processes have to
be included. The processes \gquarkS\ and \ggluonS\ contribute both to
the \gonepone-jets cross section and to the cross section for
\gtwopone-jets. In the latter case each parton in the final state
(including the photon) builds a jet on its own, whereas for
\gonepone-jets a pair of final state partons is combined into one jet.
The recombination of two partons will be performed with the help of the
cone algorithm.  The exact prescriptions will be given later when we
present our results. In the NLO calculation of the \gonepone-jet
cross section one encounters the well known infrared and collinear
singularities. For the processes \gquarkS\ and \ggluonS\ they appear in
those phase space regions where two partons are degenerate to one
parton, i.e.\ when one of the partons becomes soft or two partons become
collinear to each other. These singularities cancel in the case of soft
gluons or for collinear quark-gluon pairs against the singularities from
the virtual corrections to the LO process or have to be factorized and
absorbed into the renormalized parton distribution functions (PDF's) of
the proton. To accomplish this cancellation, the singularities are
isolated in an analytic calculation with the help of dimensional
regularization and the phase space slicing method. The technical details
of this procedure have been described in \cite{26,27} and need not to be
repeated here.

The matrix elements $|M|^2$ for the processes \gquarkS\ and \ggluonS\ 
have also photonic infrared and collinear singularities. The infrared
singularity is outside the kinematical region we are interested in since
the photon is required to be observed in the detector. In the numerical
calculation of the cross section we impose this condition by demanding a
minimal transverse momentum of the final state photon. This cut removes
also collinear singularities due to initial state photon radiation.
Final state collinear singularities are present and are treated with the
phase space slicing method similar to the case of the quark-gluon
collinear contribution. For this purpose the phase space slicing
parameter $y_0^{\gamma}$ is introduced, which is chosen very small, so
that the matrix element $|M|^2$ can be approximated by its singular
part. The slicing of the phase space is done with the help of the scaled
squared invariant masses, for example, with $y_{\gamma
  q}=(p_q+p_{\gamma})^2/W^2$ for the subprocess \gquarkS, where
$p_{\gamma}$ and $p_q$ are the four-momenta of the outgoing photon and
quark, respectively. In the gluon initiated subprocess one has two
singular regions, which are controlled by the variables $y_{\gamma q}$
and $y_{\gamma \bar{q}}$, respectively. In the region $y_{\gamma q} >
y_0^{\gamma}$ the cross section is evaluated numerically. The details
for the calculation of the various contributions are described in
\cite{26}.

The contribution to $|M|^2$ in the region $y_{\gamma q} < y_0^{\gamma}$
is collinearly divergent and is regulated by dimensional regularization.
The divergent part is absorbed into the bare photon FF to yield the
renormalized photon FF denoted by $D_{q \to \gamma}$.  The additional
fragmentation contribution to $|M|^2$ for the subprocess \gquarkS\ has
the following form
\begin{equation}
|M|^2_{\gamma^* q \rightarrow \gamma qg} 
= |M|^2_{\gamma^* q \rightarrow qg}
  \otimes D_{q \rightarrow \gamma}(z).     
\label{m2frag}
\end{equation}
There is a similar expression for the subprocess \ggluonS. It is
obvious from (\ref{m2frag}) that the fragmentation contribution is
O($\alpha \alpha_s$), the photon FF $D_{q \rightarrow \gamma}(z)$ given
by 
\begin{equation}
D_{q \rightarrow \gamma}(z)= D_{q \to \gamma}(z,\mu_{F}^2) + 
\frac{\alpha e_q^2}{2\pi}
 \left(P^{(0)}_{q\gamma}(z)\ln\frac{z(1-z)y_0^{\gamma}W^2}{\mu_F^2} 
         + z\right)\,
\label{dqg}  
\end{equation} 
being of O($\alpha$).  $D_{q \to \gamma}(z,\mu_{F}^2)$ in (\ref{dqg})
stands for the non-perturbative FF of the transition $q \to \gamma$ at
the factorization scale $\mu_{F}$, i.e.\ the scale at which the
redefinition has been performed. This function will be given by one of
the sets of photon FF mentioned above.  The second term in (\ref{dqg}),
if substituted in (\ref{m2frag}), is the finite part resulting from
adding the bare photon FF and the collinear photon-quark (-antiquark) 
contribution to the matrix element $|M|^2_{\gamma^* q \rightarrow \gamma
  qg}$ integrated in the region $y_{\gamma q} < y_0^{\gamma}$.

The $y_0^{\gamma}$ dependence in (\ref{dqg}) is expected to cancel the
dependence of the numerically evaluated \gonepone-jet cross section
restricted to the region $y_{\gamma q}>y_0^{\gamma}$ as studied in
\cite{27}. This means that $y_0^{\gamma}$ is only a technical cut
separating divergent from finite contributions.  This collinear
photon-quark (-antiquark) contribution to the matrix element
$|M|^2_{\gamma^* q \rightarrow \gamma qg}$ being calculated in the
collinear approximation is valid only up to terms O($y_0^{\gamma}$).
Consequently the cut $y_0^{\gamma}$ must be chosen sufficiently small.
The $y_0^{\gamma}$ independence must be checked by varying
$y_0^{\gamma}$.  Results of this test will be shown below.  In
(\ref{dqg}), $P^{(0)}_{q\gamma}(z)$ is the LO quark-to-photon splitting
function
\begin{equation}
     P^{(0)}_{q\gamma}(z) = \frac{1+(1-z)^2}{z}
\label{pqgamma}
\end{equation}
and $e_q$ is the electric charge of quark $q$. The variable $z$ denotes
the fraction of the final photon energy in terms of the energy of the
quark emitting the photon. If the photon is emitted from a final state
quark with four-momentum $p_{q'_4} = p_{q_4} + p_{\gamma}$, then $z$ is
obtained from $p_{\gamma}=zp_{q'_4}$.  It can also be related to the
invariants $y_{q_3\gamma}$ and $y_{q_3q_4}$ obtained from the
four-momenta of the subprocess $\gamma^*q_3 \rightarrow \gamma q_4 g$
with the result
\begin{equation}
 z = \frac{y_{q_3\gamma}}{y_{q_3q'_4}} = 
     \frac{y_{q_3\gamma}}{y_{q_3q_4}+y_{q_3\gamma}} \, .
\label{zgamma}
\end{equation}
According to (\ref{m2frag}) the fragmentation contribution to the
subprocess $\gamma^* q_3 \rightarrow \gamma q_4 g$ is calculated from
the convolution of the FF with the O($\alpha_s$) matrix element $|M|^2$
of the process $\gamma^* q_3 \rightarrow q_4 g$, which is well known. It
yields a term of O($\alpha \alpha_s$) to the cross section for
\gonepone-jets.  Similarly the fragmentation contribution to the
subprocess \ggluonS\ is calculated from the convolution of the FF for $q
\rightarrow \gamma$ with the matrix element $|M|^2$ of the process
$\gamma^* g \rightarrow q\bar{q}$ which is also known.  Equivalent
formulas are used for the calculation of the fragmentation contribution
to $\gamma^* \bar{q_3} \rightarrow \gamma \bar{q_4}g$ and \ggluonS,
where the $\bar{q}$ fragments into a photon. The fragmentation is always
assumed to occur collinearly, i.e.\ with no additional transverse
momentum in the fragmentation process.

\section{Quark-to-Photon Fragmentation Functions}
As long as we intend to calculate the \gnpone-jet cross section only up
to O($\alpha \alpha_s$) we need the photon FF only in LO, i.e.\ at
O($\alpha$).  This means that the gluon-to-photon FF, which starts at
O($\alpha \alpha_s$), is not needed. In LO the non-perturbative
quark-to-photon FF obeys the evolution equation
\begin{equation}
 \frac{dD_{q \rightarrow \gamma}(z,\mu_F)}{d\ln \mu_F^2} =
 \frac{\alpha e_q^2}{2\pi} P^{(0)}_{q\gamma}(z)
\label{dqg-evol}
\end{equation}
with the solution
\begin{equation}
D_{q \rightarrow \gamma}(z,\mu_F) = \frac{\alpha e_q^2}{2\pi}
P^{(0)}_{q\gamma}(z) \ln\left(\frac{\mu_F^2}{\mu_0^2}\right) +
   D_{q \rightarrow \gamma}(z,\mu_0) \, .
\label{dqg-solu}
\end{equation}
$D_{q \rightarrow \gamma}(z,\mu_0)$ is an initial value which must be
fitted to experimental data with a chosen initial scale $\mu_0$.  This
has been done by the ALEPH collaboration \cite{9}. From the fit to their
$e^+e^- \rightarrow \gamma + 1$-jet data \cite{9} they obtained
\begin{equation}
 D^{LO}_{q \rightarrow \gamma}(z,\mu_0)=\frac{\alpha e_q^2}{2\pi}\left(
 -P^{(0)}_{q\gamma}(z)\ln(1-z)^2 - 13.26\right)
\label{dqg-LO}
\end {equation}
with $\mu_0=0.14$ GeV. (\ref{dqg-solu}) together with (\ref{dqg-LO}) is
one of the photon FF choices which will be used to predict the cross
section $d\sigma/dz_{\gamma}$ for \gnpone-jet production in DIS
$ep$ scattering. 

We note that (\ref{dqg-solu}) is an exact solution of (\ref{dqg-evol})
at O($\alpha$). Furthermore when we substitute (\ref{dqg-solu}) into
(\ref{m2frag}) we see that together with the finite contribution
(i.e., the second term in (\ref{dqg})) the cross section becomes
independent of the factorization scale $\mu_F$. This means that 
for the cancellation of the $\mu_F$ dependence only the LO FF is needed. 
Nonetheless, in order to see the influence of the NLO corrections to
$D_{q \rightarrow \gamma}(z, \mu_F)$ we shall evaluate the \gnpone-jet
cross section also with the inclusion of the NLO photon FF.

Similarly to (\ref{dqg-solu}) the NLO FF $D_{q \rightarrow \gamma}(z,
\mu_F)$ is obtained as the solution of (\ref{dqg-evol}), but now with
O($\alpha \alpha_s$) terms added on the right-hand side of
(\ref{dqg-evol}). The result at scale $\mu_F$ is 
\begin{equation}
\begin{array}{l} \displaystyle
D_{q \rightarrow \gamma}(z,\mu_F) = \frac{\alpha e_q^2}{2\pi}
\left[P^{(0)}_{q\gamma}(z)+
      \frac{\alpha_s}{2\pi} C_F P^{(1)}_{q\gamma}(z)\right]
\ln\left(\frac{\mu_F^2} {\mu_0^2}\right) 
\\ \displaystyle
 + \frac{\alpha_s}{2\pi} C_F P^{(0)}_{qq}(z)
\ln\left(\frac{\mu_F^2}{\mu_0^2}\right) 
\otimes \left[\frac{\alpha e_q^2}{2\pi}
\frac{1}{2}P^{(0)}_{q\gamma}(z)
\ln \left(\frac{\mu_F^2}{\mu_0^2}\right) +
D_{q \rightarrow \gamma}(z,\mu_0) \right] 
   + D_{q \rightarrow \gamma}(z,\mu_0)
\end{array}
\label{dqg-solu2}
\end{equation}
where $P^{(1)}_{q\gamma}(z)$ is the next-to-leading order
quark-to-photon splitting function \cite{28} and $P^{(0)}_{qq}(z)$ is
the well-known LO $qq$ splitting function. $D_{q \rightarrow
  \gamma}(z,\mu_0)$ is the initial value of the NLO FF, which contains
all unknown long-distance contributions. The result in (\ref{dqg-solu2})
is an exact solution of the evolution equation up to O($\alpha
\alpha_s$). Based on the above arguments, also the NLO photon FF has
recently been determined \cite{10} using the ALEPH $\gamma+1$-jet data
\cite{9}. A three parameter fit with $\alpha_s=0.124$ (this value
for $\alpha_s$, just a scale-independent parameter here, 
was chosen so that the observed total $e^+e^-$ annihilation cross
section into hadrons is reproduced in the O($\alpha_s$) calculation)
yields \cite{10} 
\begin{equation}
D^{NLO}_{q \rightarrow \gamma}(z,\mu_{0})=\frac{\alpha e_q^2}{2\pi} \left(
-P^{(0)}_{q\gamma}(z) \ln(1-z)^2 + 20.8(1-z) - 11.07 \right)
\label{dqg-NLO}
\end{equation}
with $\mu_0 = 0.64$ GeV. Inside the experimental errors this fit for the
photon FF at $\mu_{0}$ describes the ALEPH data as good as the LO fit
(\ref{dqg-LO}) \cite{10,11}. A fit with a larger value of
$\alpha_s$ is also reported in \cite{11}. 

It should be noted that the LO and NLO FF's of the photon as given in
(\ref{dqg-solu}) together with (\ref{dqg-LO}) and in (\ref{dqg-solu2})
together with (\ref{dqg-NLO}) do not take into account resummation of
powers of $\ln (\mu_F^2/\mu_0^2)$ as usually implemented, e.g.\ via
Altarelli-Parisi evolution \cite{29}. Such resummations are only
unambiguous if the resummed logarithm is the only large logarithm in the
kinematical region under consideration. If logarithms of different
arguments can become simultaneously large, the resummation of one of
these logarithms at a given order implies that all other potentially
large logarithms are shifted into a higher order of the perturbative
expansion, i.e.\ are neglected. In the evaluation of the $\gamma+1$-jet
rate at $O(\alpha)$ \cite{11} and $O(\alpha \alpha_{s})$ \cite{8} at
LEP for $0.7 < z < 1$, one encounters at least two different potentially
large logarithms, $\ln \mu^2_{F}$  and $\ln (1-z)$. In the high $z$
region, the region where the photon is isolated or almost isolated it is
by far not clear that $\ln \mu^2_{F}$ is the largest logarithm. Choosing 
not to resum the logarithms of $\ln \mu_F$ is therefore equally
justified for the case of large $z \rightarrow 1$.

In the conventional approach, the parton-to-photon FF's $D_{i
 \rightarrow \gamma}(z,\mu_F)$ satisfy a system of inhomogeneous
evolution equations \cite{29}. Usually these equations are diagonalized
in terms of the singlet and non-singlet quark FF's as well as the gluon
FF. For the case that the gluon-to-photon fragmentation is neglected,
which we shall do in our application to DIS $\gamma$+jets production,
these equations can be simplified \cite{12}. Then the flavor singlet
and non-singlet quark-to-photon FF's become equal to a unique function
$D_{q \rightarrow \gamma}$. This function satisfies the
all-orders evolution equation 
\begin{equation}
\frac{dD_{q \rightarrow \gamma}(z,\mu_F)}{d\ln \mu_F^2} =
\frac{\alpha e_q^2}{2\pi}P^{(0)}_{q\gamma} 
+ \frac{\alpha_s(\mu_F)}{2\pi}
  P^{(0)}_{qq}(z) \otimes D_{q \rightarrow \gamma}(z,\mu_F) 
\label{dqg-evol-in}
\end{equation}
which has a similar form as the next-to-leading order evolution in the
fixed-order approach, but the coupling $\alpha_s$ is not fixed and now
is taken as a function of the factorization scale $\mu_F$. The full
solution $D_{q \rightarrow \gamma}(z,\mu_F)$ of the all-orders
inhomogeneous evolution equation is a sum of two contributions. The
first term is the point-like (or perturbative) part $D^{pl}_{q
  \rightarrow \gamma}(z,\mu_F)$, which in the leading logarithmic (LL)
approximation is a solution of the inhomogeneous equation
(\ref{dqg-evol-in}). The second term is the hadronic (or
non-perturbative) part $D^{had}_{q \rightarrow \gamma}(z,\mu_F)$, which
is the solution of the corresponding homogeneous equation. This term
must be fitted to experimental data to obtain the input at a starting
scale. At LL only terms of the form $\alpha_s^n \ln^{n+1}\mu_F^2$ are
kept while beyond leading logarithms (BLL) both the leading and the
sub-leading ($\alpha_s^n \ln^n\mu_F^2$) logarithms of $\mu_F$ are
resummed to all orders in the strong coupling constant $\alpha_s$. Thus
the LO solution (\ref{dqg-solu}) is the $n=0$ term of the LL
approximation, whereas the NLO solution (\ref{dqg-solu2}) is the BLL
approximation up to the order $n=1$. By dropping the $\alpha_s
\ln\mu_F^2$ terms in (\ref{dqg-solu2}) one can also infer the LL
approximation up to the order $n=1$. 

In this conventional approach, the quark-to-photon fragmentation
function is regarded as being of $ O(\alpha/\alpha_{s})$. In fact, the
$n=0$ term of the LL approximation is proportional to
$\ln(\mu_F^2/\mu_0^2)$. At large scales $\mu_F^2 \gg \mu_0^2$ and
assuming that $\mu_F$ can be identified with the scale that determines
$\alpha_s$, i.e.\ $\alpha_s \sim 1/\ln \mu_F^2$, the LL FF becomes of
the order of O($\alpha/\alpha_s$). This motivates the usual statement
that the total fragmentation contribution is like O($\alpha$), i.e.\ of
the same order as the LO direct photon contribution. Of course, this
statement depends on the considered kinematic region. With this argument
however, one usually justifies the inclusion of terms in the
fragmentation contribution of O($\alpha \alpha_s^2$) which become of
O($\alpha \alpha_s$) if the FF is considered of O($\alpha/\alpha_s$). 

The most recent parametrizations of photon FF's in the conventional
approach as described in the previous paragraph are those of GRV
\cite{4} and BFG \cite{5}. The FF's of GRV
\cite{4} exist in the LL as well as in the BLL form, which in the
following we shall denote as the LO and NLO FF's of GRV. However, the FF
of BFG \cite{5} are available only in the BLL approximation. They both
have been compared to the ALEPH $\gamma +1$-jet cross section \cite{9},
which is sensitive to the large $z$ region ($0.7<z<1.0$). The
BFG prediction goes through the experimental points whereas the GRV NLO
parametrization lies systematically higher and agrees less well with the
ALEPH data \cite{12,13}. This difference can be attributed to both the
choice of the input scale $\mu_0$ and of the non-perturbative input at
this scale. The BFG input is smaller and according to the ALEPH data this
choice is preferred. 

Thus we have at least five different versions of photon FF's at our
disposal which have been compared to the ALEPH data: two LO
parametrizations, the one using the ALEPH data directly to determine the
initial distribution given in (\ref{dqg-solu}) and (\ref{dqg-LO}) and
the GRV parametrization in LO; and three NLO choices, the one written in
(\ref{dqg-solu2}) and (\ref{dqg-NLO}) directly fitted to the ALEPH data,
and the NLO parametrizations of GRV and BFG. We shall use these five
parametrizations to predict the \gnpone-jet cross section in
deep-inelastic $ep$ scattering. 

\section{Numerical Results}
The results for these cross sections are obtained for energies and
kinematical cuts appropriate for the HERA experiments. The energies of
the incoming electron and proton are $E_e=27.5$ GeV and $E_p=820$ GeV,
respectively. The cuts on the DIS variables are as in our previous work
\cite{26} 
\begin{eqnarray}
   Q^2 \geq 10~{\rm GeV}^2,~~~~~W \geq 10~{\rm GeV}\, ,  
\nonumber \\
   10^{-4} \leq x \leq 0.5,~~~~~~0.05 \leq y \leq 0.95\, .
\label{cuts1}
\end{eqnarray}
To reduce the background from lepton radiation \cite{27} we require
\begin{equation}
90^{\circ} \leq \theta_{\gamma} \leq 173^{\circ},~~~~~~
\theta_{\gamma e} \geq 10^{\circ}
\label{cuts2}
\end{equation}
where $\theta_{\gamma}$ is the emission angle of the photon measured
with respect to the momentum of the incoming electron in the HERA
laboratory frame. The cut on $\theta_{\gamma e}$, the angle between the
photon and the outgoing electron momentum, suppresses radiation from the
final-state electron. The $\gamma $ and the hadron jet $J$ are required
to have minimal transverse momenta
\begin{equation}
 p_{T,\gamma} \geq 5~{\rm GeV},~~~~~~~~~p_{T,J} \geq 6~{\rm GeV}\, .
\label{cuts-pt}
\end{equation}
Different values of minimal $p_T$'s for the photon and the jet have to
be chosen in order to avoid the otherwise present infrared sensitivity
of the NLO predictions \cite{26}. The PDF's of the proton are taken from
MRST \cite{30}. $\alpha_s$ is calculated from the two-loop formula with
the same $\Lambda $ value ($\Lambda_{\overline{MS}}(n_f=4)=300$ MeV) as
used in the MRST parametrization. The scale in $\alpha_s$ and the
factorization scales $\mu_F$ for the proton PDF and the photon FF are
equal and fixed to $\sqrt{Q^2}$. 

We are interested in the differential cross section
$d\sigma/dz_{\gamma}$ at NLO (up to O($\alpha \alpha_s$)) as a function
of $z_{\gamma}$, where 
\begin{equation}
 z_{\gamma } = \frac{p_{T,\gamma}}{p_{T,\gamma}+p_{T,had}}\, .
\label{zgamma-pt}
\end{equation}
This definition of $z_{\gamma}$ agrees with the one in (\ref{zgamma})
for the fragmentation contribution. In (\ref{zgamma-pt}), $p_{T,had}$ is
the transverse momentum of the parton producing hadrons, which is
recombined with the photon into the photon jet. Here the photon is
treated like any other parton during recombination, so that one of the
recombined jets may be the photon jet. For the recombination of the two
partons into a hadron jet or a parton and the photon into the photon
jet, we use the cone algorithm of the Snowmass convention \cite{31}. The
recombination is applied in the $\gamma +2$ parton sample and yields a
contribution to the \gonepone-jet class. The \gtwopone-jet class
consists of the unrecombined contributions of the $\gamma +2$ parton
sample. The recombination is performed in the $\gamma^* p$
center-of-mass frame. Two partons $i$ and $j$ are combined into a jet
$J$ if they obey the cone constrains $R_{i,J}<R$ and $R_{j,J}<R$, where 
\begin{equation}
 R_{i,J} =\sqrt{(\eta_i-\eta_J)^2 + (\phi_i-\phi_J)^2}\, .
\label{Rij}
\end{equation}
$\eta_J$ and $\phi_J$ are the rapidity and the azimuthal angle of the  
recombined jet. If for example, parton $i$ is the $\gamma $, then $J$ is
the photon jet. The jet variables $\eta_J$ and $\phi_J$ are obtained
from the averages of the corresponding variables of the recombined
partons $i$ and $j$ multiplied with their respective $p_T$ values and
$p_{T,J}=p_{T,i}+p_{T,j}$. We choose $R=1$. The azimuthal angle is
defined with respect to the scattering plane defined by the momenta of
the ingoing and outgoing electron. It is known that the cone algorithm
is ambiguous for final states with more than three particles or
partons. Since we have maximally three partons in the final state this
is not relevant in our case. Furthermore it will be no problem to repeat
the calculation for any other cluster algorithm that might be used in
the analysis of forthcoming experimental data. It is obvious that for
the fragmentation contribution the cone constraint is always satisfied
since the hadronic remnant in the fragmentation is collinear with the
photon.  

\begin{figure}[tb] 
\unitlength 1mm
\begin{picture}(160,100)
\put(30,-1){\epsfig{file=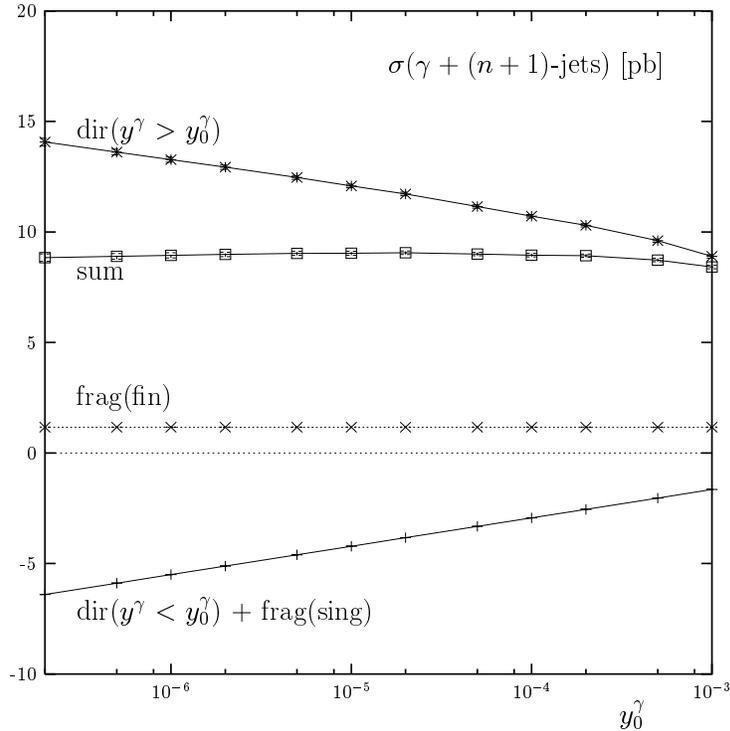,height=10cm,width=10cm,%
                   bbllx=66,bblly=250,bburx=510,bbury=710,clip=}}
\end{picture}
\caption{$y_0^{\gamma}$-dependence of separate contributions to the
  \gnpone-jets cross section.}
\label{fig1}
\end{figure}

Before we present our results for $d\sigma/dz_{\gamma}$ for the various 
choices of the photon FF we show that the cross section is independent
of the slicing cut $y_0^{\gamma}$. For this purpose we have calculated
the sum of the cross sections for the production of \gnpone-jets
$(n=1,2)$ as a function of $y_0^{\gamma}$ in the range $2\cdot
10^{-7}<y_0^{\gamma}<10^{-3}$. The result is shown in Fig.\
\ref{fig1}. Here we have plotted the cross section for the direct
contribution with $y_{\gamma}>y_0^{\gamma}$ which increases with
decreasing $y_0^{\gamma}$ and the direct contribution for 
$y_{\gamma}<y_0^{\gamma}$ together with the second term in contribution
(\ref{dqg}) inserted in (\ref{m2frag}) which we denote by 'frag(sing)'
in the figure. This latter contribution is negative and decreases with
decreasing $y_0^{\gamma}$ in such a way that the sum of the increasing
and decreasing contributions is constant inside the numerical accuracy
for $y_0^{\gamma} < 5\cdot 10^{-4}$. The sum shown in Fig.\ \ref{fig1}
contains also the fragmentation contribution (\ref{m2frag}) with the
photon FF $D_{q \to \gamma}(z,\mu_{F}^2)$ as given in (\ref{dqg-solu})
and (\ref{dqg-LO}) (denoted by 'frag(fin)' in the figure). Of course,
this contribution is independent of $y_0^{\gamma}$. It amounts to
approximately $14\%$ of the total \gnpone-jet cross section. The cross
sections shown in Fig.\ \ref{fig1} are calculated with the kinematical
cuts specified above and the additional cut $z_{\gamma} \geq 0.5$. From
Fig.\ \ref{fig1} we conclude that the slicing cut for the photon should
be chosen smaller than $10^{-4}$ to obtain a reliable cross section. For
the following results we fixed it to $y_0^{\gamma} = 10^{-5}$ as in our
previous work \cite{26}.  

\begin{figure}[bt] 
\unitlength 1mm
\begin{picture}(160,100)
\put(30,-1){\epsfig{file=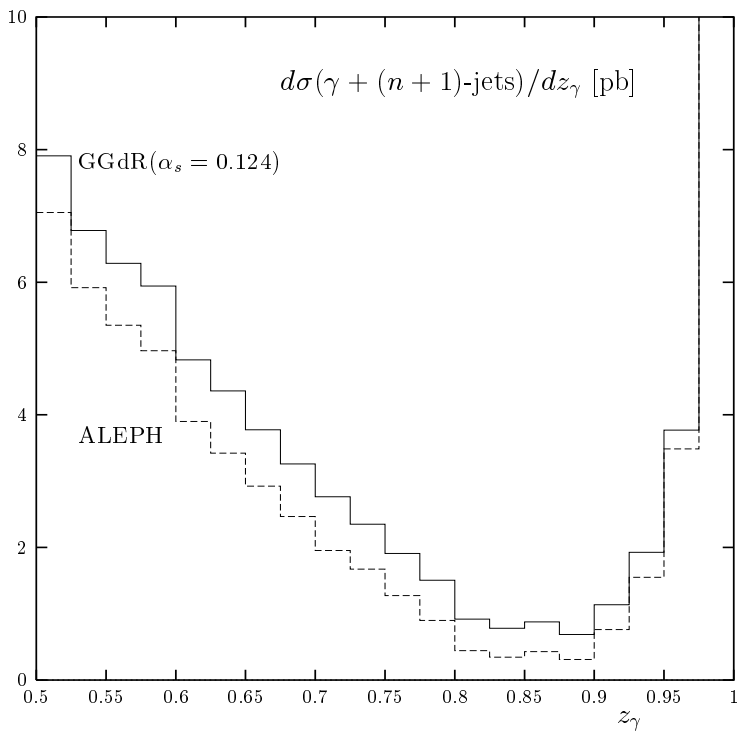,height=10cm,width=10cm,%
                   bbllx=50,bblly=250,bburx=480,bbury=710,clip=}}
\end{picture}
\caption{Comparison of the \gnpone-jet cross section for the LO and NLO
  parametrizations of the FF obtained from fits to the ALEPH data.}
\label{fig2}
\end{figure}

\begin{figure}[tb] 
\unitlength 1mm
\begin{picture}(160,100)
\put(30,-1){\epsfig{file=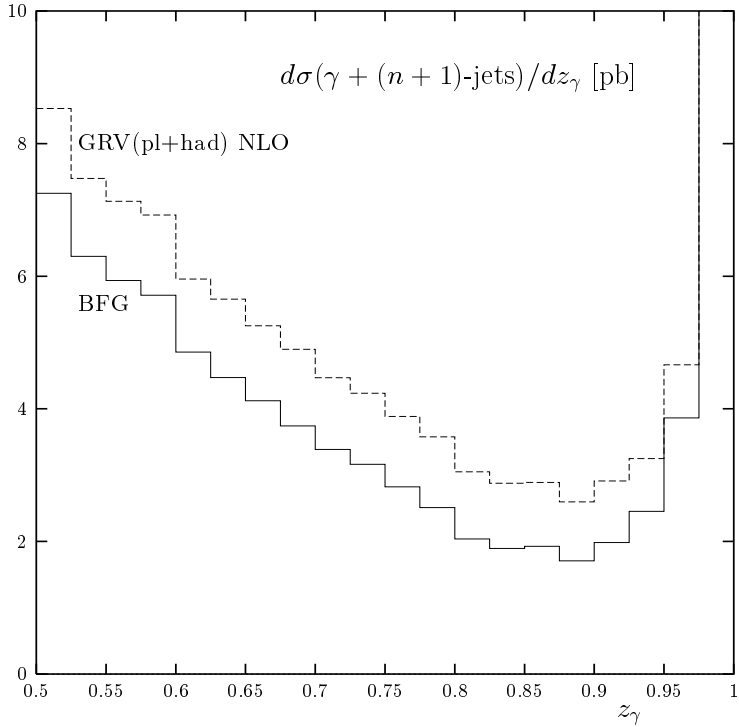,height=10cm,width=10cm,%
                   bbllx=50,bblly=250,bburx=480,bbury=710,clip=}}
\end{picture}
\caption{Comparison of the \gnpone-jet cross section for the NLO
  parametrizations of GRV and BFG.}
\label{fig3}
\end{figure}

\begin{figure}[tb] 
\unitlength 1mm
\begin{picture}(160,100)
\put(30,-1){\epsfig{file=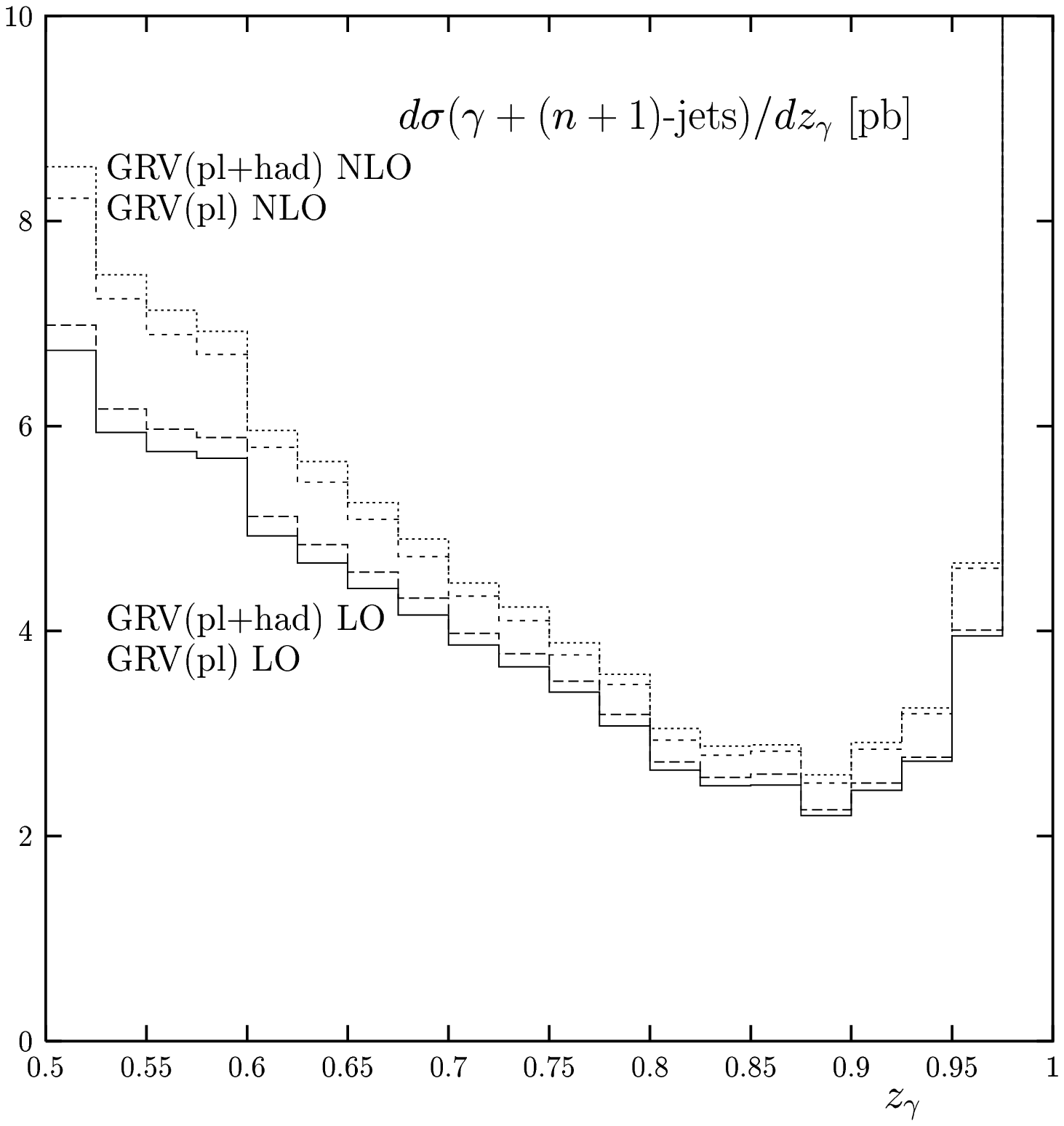,height=10cm,width=10cm,%
                   bbllx=50,bblly=250,bburx=480,bbury=710,clip=}}
\end{picture}
\caption{Comparison of the \gnpone-jet cross section for LO and NLO
  parametrizations of GRV.}  
\label{fig4}
\end{figure}

The results for $d\sigma/dz_{\gamma}$ as a function of $z_{\gamma}$ are 
shown in Figs.\ \ref{fig2}, \ref{fig3} and \ref{fig4}. The total cross
section is dominated by the last bin $0.975 < z_{\gamma} \le 1$ which is 
clipped off above 10 pb in these figures. In this bin we have the direct
contribution and those fragmentation contributions with no or little
hadronic remnant. The fraction of the total cross section (for $0.5 \leq
z_{\gamma} \leq 1$) contained in the last bin amounts to $70\,\%$ to
$80\,\%$ and does not depend on the parametrization of the FF. Therefore
it is clear that the dependence of the photon plus jet cross section on
$z_{\gamma}$ below $z_{\gamma} = 1$ is an appropriate observable which
contains information on the photon FF. Here the fragmentation
contribution dominates and in addition we have those direct
contributions from the ($\gamma + 2$)-parton sample where the photon is
recombined with one of the partons. In Fig.\ \ref{fig2} the cross
section $d\sigma/dz_{\gamma}$ in the region $0.5 < z_{\gamma} < 1.0$ is
plotted for the photon FF's from (\ref{dqg-solu}) and (\ref{dqg-LO})
(denoted ALEPH) and from (\ref{dqg-solu2}) and (\ref{dqg-NLO}) (denoted
GGdR($\alpha_s=0.124$) in the figure). For both cases the cross section
first decreases with increasing $z_{\gamma}$ until it reaches a minimum
near $z_{\gamma} = 0.9$, from which it increases strongly towards
$z_{\gamma} = 1$. This dependence is similar to what has been
found for the photon plus one-jet cross section in $e^+e^-$ annihilation
at the $Z$-resonance \cite{10,11}. Note however, that below
$z_{\gamma}=0.5$, i.e.\ outside the range shown in these figures, the
cross section decreases again towards smaller values of $z_{\gamma}$.
Unlike at LEP, the photon-jet is here required to have a minimal
transverse momentum as given by (\ref{zgamma-pt}). The decrease in the
cross section for values of $z_{\gamma}$ below 0.5 is a kinematical
effect resulting from the imposition of this cut. 

For the two choices of FF's the predicted cross section differs by
approximately $20 \%$. The cross section for the LO FF is smaller than
for the NLO FF. We note that the $\alpha_s$ value quoted for this FF is
not the value for which the cross section was calculated; rather it is
the value which was chosen to fit the photon FF to the ALEPH data. Both
photon FF's fit the ALEPH data in LO and NLO approximation. In our case
only the LO FF is appropriate. The result for the NLO FF gives just an
indication of higher-order corrections up to O($\alpha \alpha_s^2$). 
However, it is not a full prediction at this order since the NNLO
calculation for the direct contribution has not been done, the NLO
corrections to $|M|^2$ on the right-hand-side of (\ref{m2frag}) are not
included either.    
%

In Fig.\ \ref{fig3} the predictions with GRV \cite{4} and BFG \cite{5}
FF's are shown. Both are FF's in NLO and with all leading and subleading
logarithms of $\ln(\mu_F^2/m_0^2)$ resummed at the next-to-leading 
logarithmic accuracy. The behavior of the cross section as a function
of $z_{\gamma}$ is qualitatively similar to the result in Fig.\
\ref{fig2}. The cross section for GRV is larger than the one with the
BFG FF. This is to be expected since GRV gives a larger $e^+e^-$ cross
section for $(\gamma+1)$-jet than BFG also for the ALEPH kinematical
conditions.  

Comparing with the fixed-order result in Fig.\ \ref{fig2} the BFG cross
section lies above the fixed-order prediction obtained using the LO FF
over the whole $z_{\gamma}$ region. The discrepancy is largest in the
region $0.65<z_{\gamma}<0.95$ where the cross section is smaller. At the
minimum the two cross sections differ by more than a factor two. 

In Fig.\ \ref{fig4} the results for the GRV photon FF's are displayed,
but now for both the LO and the NLO parametrization. On the average the
cross section for the LO fragmentation function is again smaller by
$20\%$ compared to the NLO prediction. In addition, we have  plotted the
cross sections for the point-like approximations of the FF's. This
approximation is rather good, i.e.\ the influence of the hadronic part
of the FF is small in the considered $z_{\gamma}$ range. We note that
the difference between the full FF and the point-like part increases
with decreasing $z_{\gamma}$. 

If we compare our results for the various photon FF's in Figs.\
\ref{fig2}, \ref{fig3} and \ref{fig4} we observe that the predictions
agree approximately within $20\%$ in the small and large $z_{\gamma}$
regions, i.e.\ for $z_{\gamma} <0.65$ and $z_{\gamma}>0.90$. However,
near the minimum of the cross section, i.e.\ in the region
$0.65<z_{\gamma}<0.95$, the results differ by up to a factor two. The
largest differences occur between the predictions obtained with the
ALEPH photon fragmentation function on the one side and the GRV and BFG
parametrizations on the other side. This difference comes mainly from
the fact that different evolution approaches are used. Whereas for GRV
and BFG the FF at $\mu_F^2=Q^2$ is obtained from the conventional
evolution after the leading and/or subleading logarithms of $\mu_{F}$
were resummed, the ALEPH FF's are evolved only to the respective finite
order in $\alpha_{s}$ as given in (\ref{dqg-solu}) and
(\ref{dqg-solu2}). Therefore, if we calculated $d\sigma/dz_{\gamma}$ for 
$ep \rightarrow e\gamma + (n+1)$-jets at the large scale $\mu_F=M_Z$
the cross sections obtained for BFG and ALEPH would come out quite
similar over the whole $z_{\gamma}$ range inside the $20\%$ margin. Only
when we go to the scale $\mu_F^2=Q^2$, which on the average is much
smaller, we observe that the cross section obtained using the BFG photon
fragmentation function is much larger than the ALEPH cross section in
the region $0.65 < z_{\gamma} <0.95$. 

Provided the resummed solution of the all-orders evolution equation can
be accurately determined \cite{12} over the whole $z_{\gamma}$ range under
consideration, i.e.\ for $0.5<z_{\gamma}<1$, the approach using this
solution represents the theoretically preferred approach as it is the
most complete. The approach using an expanded and therefore approximated
photon FF has however important advantages. As already mentioned, its
use leads to factorization scale independent results for the cross
section evaluated at a given fixed order in $\alpha_{s}$. Moreover it
enables an analytic determination of the photon FF. 

Within the conventional approaches, the BFG prediction should be
preferred over the result with the GRV FF. The difference between GRV
and BFG in Fig.\ \ref{fig3} is related to the choice of a different
input at the starting scale $\mu_0$. As already mentioned, the BFG
parametrization is in better agreement with the ALEPH data \cite{12}
than the GRV NLO parametrization and therefore is the more realistic
choice of FF for the photon. This could be further tested as soon as
data from HERA on photon plus jet production in DIS become available. 

\section{Conclusions}
We have presented a NLO calculation for the production of photons
accompanied by jets in deep inelastic electron proton scattering,
taking into account the contribution from quark-to-photon
fragmentation. We have calculated the cross section as a function of
$z_{\gamma}$ for those parametrizations of the photon fragmentation
functions, which had been compared to photon plus jet data from the
ALEPH collaboration. As we observed significant differences between the
predictions, we expect that the measurement of photon plus jet 
production in DIS at HERA will contribute to testing these photon
fragmentation functions. 

\subsection*{Acknowledgements}

A.\ G.\ would like to thank A.\ Wagner for financial support
during her stay at DESY where part of this work has been carried out. 


\end{document}